# The densification mechanism of squeeze casting

Xiao-Hui Chen[1], Xu Huang[2], Xue-Ping. Ren[1*]

[1] School of Materials Science and Engineering，University of Science and Technology Beijing，Beijing 100083，P.R. China

[2] Department of Mechanical Engineering, Yale University, New Haven, Connecticut 06520, USA

## Abstract

The solidification process during squeeze casting is analyzed based on the classical solidification and plastic deformation theory. The linear relationship between punch velocity and the solidification rate is established if the density change of molten meals is neglected. To obtain defect-free castings, the punch velocity should be larger than the solidification rate. The densification mechanism is also discussed. The plastic deformation will result in the radical movement of the central molten metals, which reduces the temperature gradient from the center to the mold wall, which provides the condition to obtain defect-free castings: simultaneous nucleation.

*Correspondence author; email address: rxp33@mater.ustb.edu.cn

## 1. Introduction

Squeeze casting technology has been widely investigated since cast products with less defect can be obtained [1-3]. Comparing to the conventional casting method, the melting point of metal is increased due to the applied stress, implying a higher supercooling degree[4,5]. Additionally, the effect contact area between the mold wall and the casting products increases, leading to a higher heat transfer coefficient and solidification rate. As such, the whole solidification time is decreased during squeeze casting [6-8]. However, it is difficult to determine the solidification rate during conventional and squeeze casting from both theoretical and practical operations. The purpose of the present study is to establish the relationship between the punch velocity and the solidification rate during squeeze casting, and the obtained relationship is verified experimentally. On the other hand, it is impossible to obtain totally defect-free casting product based on the solidification and plastic deformation models. Thus, the condition to obtain defect-free casting products is further discussed in the present study. This work will shed some light on the applications of squeeze casting in industry.

## 2. The squeeze casting process

### The relationship between the punch velocity and solidification rate

Squeeze casting is a technique to control the metal solidification process by applied pressure, and the whole solidification is processed in a close mold. The molten metals are first poured into the mold, and then an extra pressure is applied on them through a punch. The squeeze casting process is schematically shown in Figure 1a. The whole casting metal is composed of the solidified component, the molten metals, and a small coexist region between them. The solidified component is usually at the outside of the casting metals, named solid shell metals in the following. The coexist-region is neglected during the following analysis for simplification. As shown in Figure 1a, a cavity is formed due to the solidification shrinkage during the squeeze casting. The condition to eliminate this shrinkage cavity is that there is a plastic deformation of the solid shell metals during squeeze casting. So, the applied stress should be above the yield stress of the examined metal, which is shown in Fig. 1b.

The volume change can be neglected during plastic deformation and the cooling process during squeeze casting. So, the punch velocity $\dot{u}$ can be expressed as,

$$\dot{u} = \frac{du}{dt} \quad (1)$$

where $u$ is displacement of the punch, and $t$ the time.

The solidification rate of molten metals $\dot{v}$ is given by,

$$\dot{v} = \frac{dV}{dt} \quad (2)$$

where $dV$ is the volume change due to solidification during the time *dt*.

During the time segment *dt*, the total punch movement volume change $dV_0$ is composed of the volume shrinkage of molten metals $dV_d$ due to the applied pressure $P$ and the solidification volume change. The solidification volume change can be expressed as the product of volume change per molar $\Delta V$ and the involved volume $dV$. So,

$$dV_0 = A\dot{u}dt = \Delta V dV + dV_d \quad (3)$$

A is the contact area between the metals and the punch.

Based on the mass conservation condition,

$$m = \rho \cdot V_l = C \quad (4)$$

where *m* is the mass of molten metals, *ρ* is the density of the molten metals, and $V_l$ is the volume of the molten metals. So, the volume shrinkage of molten metal $dV_d$ due to the applied pressure *P* can be expressed as

$$dV_d = V_d \frac{d\rho}{\rho} \quad (5)$$

Substitute Eq. (2) and (5) into Eq.(3), the punch velocity $\dot{u}$ can be expressed as,

$$\dot{u} = \frac{\Delta V}{A}\dot{v} + \frac{V_d}{\rho \cdot A} \cdot \frac{d\rho}{dt} \quad (6)$$

The density of molten metals $\rho$ is a function of applied pressure *P*. Thus,

$$\frac{d\rho}{dt} = \frac{d\rho}{dP} \cdot \frac{dP}{dt} \quad (7)$$

Substituting Eq.(7) into Eq.(6), the punch velocity $\dot{u}$ can be expressed as a function of the solidification rate of molten metals $\dot{v}$ and the applied pressure *P*.

$$\dot{u} = \frac{\Delta V}{A}\dot{v} + \frac{V_d}{\rho \cdot A} \cdot \frac{d\rho}{dP} \cdot \frac{dP}{dt} \quad (8)$$

The punch velocity $\dot{u}$ is closely related to the plastic deformation behavior of the solid shell metals. Thus, the relationship between plastic deformation and solidification behavior was established through Eq.(8) during squeeze casting.

When the applied pressure is above the critical value, the density change of molten metals with the applied pressure can be neglected. So Eq.(8) can be further simplified as

$$\dot{v} = \frac{A}{\Delta V}\dot{u} \quad (9)$$

The solidification rate of molten metals $\dot{v}$ is a function of the punch velocity $\dot{u}$, which can be determined experimentally. As such, the solidification rate of molten metals $\dot{v}$ can be obtained through Eq. (9) although it is well known that the solidification rate is very difficult to measure directly during squeeze casting.

**The minimum requirement for punch velocity**

The punch velocity is required to reach a certain value to obtain high quality casting products during squeeze casting. It can be classified into three-case based on the maximum punch velocity $\dot{u}_{max}$. Following Eq. (9),

(1) If $\dot{u}_{max} < \frac{\Delta V}{A}\dot{v}$

In this case, there is no any pressure applied on the central molten metals during solidification, leading to the shrinkage cavities in the central of the casting components. This solidification process is similar with that of the die-casting. Based on the plastic deformation theory of metals, infinite big pressure is required to eliminate the shrinkage cavities after casting process.

(2) If $\dot{u}_{\max} = \frac{\Delta V}{A}\dot{v}$

The condition of Eq. (9) is satisfied in this case, and the punch speed is compatible with the solidification rate of molten metals. The subjected pressure of the central molten metals, however, depends on the initial applied pressure. When the initial applied pressure is low, it is similar as case (1), i.e., the shrinkage cavities can be formed in the central of the casting products. On the other hand, defect-free casting products can be obtained when the initial applied pressure is high.

(3) If $\dot{u}_{\max} > \frac{\Delta V}{A}\dot{v}$

In this case, the central molten metals are always under the isostatic stress. Additionally, the isostatic stress is almost equal to the applied pressure, which increases the density of the molten metals. In other words, the squeeze casting condition is satisfied. When the density of the molten metals is up to that of the solid one, defect-free casting product can be obtained.

**The experimental method and results**

A 6063 Al alloy was used to evaluate the above theory, whose chemical composition is shown in Table 1. The inner diameter of the mold is 50 mm, which was machined using 3Cr2W8V tool steel. Water-base graphite is used as the lubricant, and the mold was pre-heated to 280$^o$C. The Al alloys were molten in crucible, and the pour temperature is 700℃ and the height of pouring about 35 mm. The experiment was processed in a LB25A hydraulic machine, and the applied pressure is selected as 49.5MPa and 74MPa. The displacement of punch movement was measured using a displacement sensor, and the punch speed was calculated using the measured displacement.

The working pressure fluctuates with time during squeeze casting is shown in Figure 2. In both applied pressure, the solidification rate of molten metals is smaller than the maximum punch velocity. When the applied pressure is 49.5 MPa, it is saturated immediately and keeps constant during the whole squeeze casting process. When the applied pressure is increased to 74 MPa, however, the pressure smoothly goes to that value in about 15 seconds, which is related to the densification of the molten metals.

The displacement of punch movement as a function of time is shown in Figure 3. The displacement has some fluctuation during the squeeze casting, indicating that it depends on the plastic deformation of the solidified shell metals. When the punch speed cannot keep pace with the solidification rate, the fluctuation occurs. The applied pressure increases the solidification rate of molten metals, and 5 seconds are enough to fully solidify the alloys when the applied pressure increases from 49.5MPa to 74MPa in the present study.

The relationship between the punch velocity and time can be obtained from Figure 3, which is shown in Figure 4. The fluctuation amplitude increases with the applied pressure. This is related to the higher solidification rate under higher applied pressure, and the un-uniform plastic deformation of the solidified shell metals. All these lead to the solidification rate differs from the punch speed. Based on the smoothly fitting line in Figure 4, the solidification rate of 6063 Al alloy can be obtained.

## 3. The densification mechanism during squeeze casting

The squeeze casting is usually processed in a closed mold, and the last region to solidify is in the central of the processing component based on the classic solidification theory [9]. On the other hand, it is impossible to eliminate the shrinkage void due to solidification in a closed mold based on the plastic deformation theory [10]. To solve this discrepancy, the densification mechanism during squeeze casting under ideal solidification condition is briefly discussed in the following.

### The densification mechanism under ideal conditions

The ideal solidification condition of metals indicates that the molten component will always be uniform during the whole casting process. In other words, the molten metals will be simultaneously nucleated.

According to the Clasius-Clapeyron equation [11], the melting point of metals and alloys increases with applied pressure.

$$T = T_R \exp\left(\frac{P\Delta V}{\Delta H}\right) \tag{10}$$

where $T_R$ is the melting point of metals and alloys under 1MPa pressure. $P$ the applied pressure, $\Delta H$ the latent heat of melting per molar, and $\Delta V$ the molar volume change during solidification.

If the pouring temperature is low, the simultaneously nucleation can be achieved under very low pressure, which results in a defect-free casting products.

**The densification mechanism under actual condition**

The molten metals and alloys cannot be simultaneously nucleated during the practical squeeze casting. There is a temperature gradient from the center to the wall of the mold. During squeeze casting, the solidification starts from the mold wall due to the high coefficient of heat transfers and larger supercooling degree, and the solidification will proceed to the center gradually. Solidification shrinkage occurs due to the volume change, which provides space for plastic deformation of the solid shell metals. In turn, this plastic deformation makes the molten metals move radically, which decreases the temperature gradient.

Supposed the inner diameter of the cast mold is $D$, the last solidification region is a cylinder with a diameter $d$, and the applied pressure is $P_0$, then the pressure on the central molten metals $P$ is,

$$P = \frac{D^2}{d^2} P_0 \tag{11}$$

Obviously, $P > P_0$ since $D > d$. With the solidification processing during squeeze casting, the difference between $D$ and $d$ increases, resulting in a much larger applied pressure on the central molten metals. This is the reason that densification and defect-free casting products can be obtained during squeeze casting.

In summary, both the radical movement and the larger pressure of central molten metals are contributed to the densification and defect-free products from squeeze casting.

### The coefficient of simultaneous solidification

Supposed the volume of molten metals is $V_0$, and the volume of metals that can be nucleated simultaneously is $V'$, the coefficient of simultaneous solidification $\lambda$ is,

$$\lambda = \frac{V'}{V_0} \qquad (12)$$

The coefficient $\lambda$ reflects the volume of the molten metals in the central region during squeeze casting. The condition of $\lambda = 1$ indicates the ideal condition of squeeze casting, and that of $\lambda = 0$ the traditional die-casting, where the micro-void cannot be totally eliminated. The larger $\lambda$ indicates that the volume of last solidification region is large, and the volume of the solidification shell is small. So the required application pressure is low, and the duration for solidification is short.

## 4. Conclusion

(1) The linear relationship between the punch velocity and the solidification rate is obtained if the density change during solidification is neglected. So the solidification rate can be easily obtained from the measured punch velocity from Eq. (9).

(2) The condition to obtain densification and defect-free casting product is that the punch velocity is larger than that of the solidification rate during squeeze casting.

(3) The radical movement of the molten metals during squeeze casting reduces the temperature gradient, which makes the simultaneous nucleation possible.

## Table and Figures Captions

| | |
|---|---|
| Table 1 | The chemical composition of 6063 Al alloys (wt.%) |
| Figure 1 | Schematically show the solidification process during squeeze casting. |
| Figure 2 | The working pressure fluctuates with time for a 6063 Al alloy during squeeze casing. |
| Figure 3 | The displacement of the punch as a function of time for a 6063 Al alloy during squeeze casing. |
| Figure 4 | The punch speed change with time for a 6063 Al alloy during squeeze casing. |

Table 1 The chemical composition of 6063 Al alloys (wt.%)

| Si | Fe | Cu | Mn | Mg | Cr | Zn | Ti | Al |
|---|---|---|---|---|---|---|---|---|
| 0.49 | 0.31 | 0.10 | 0.10 | 0.61 | 0.06 | 0.03 | 0.03 | Residuals |

Figure 1

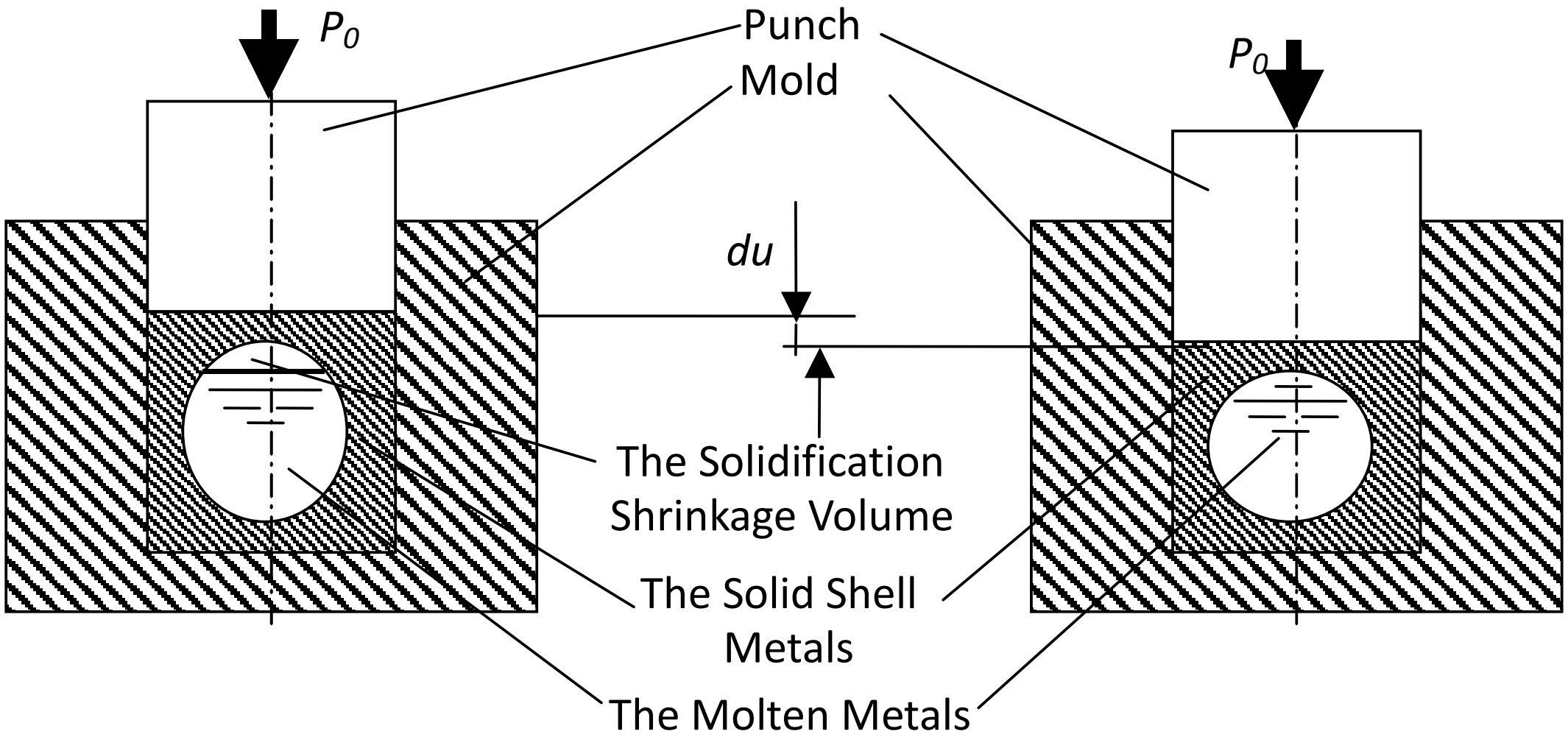

(a) The Squeeze Casting Process

(b) During the time segment *dt*

Figure 2

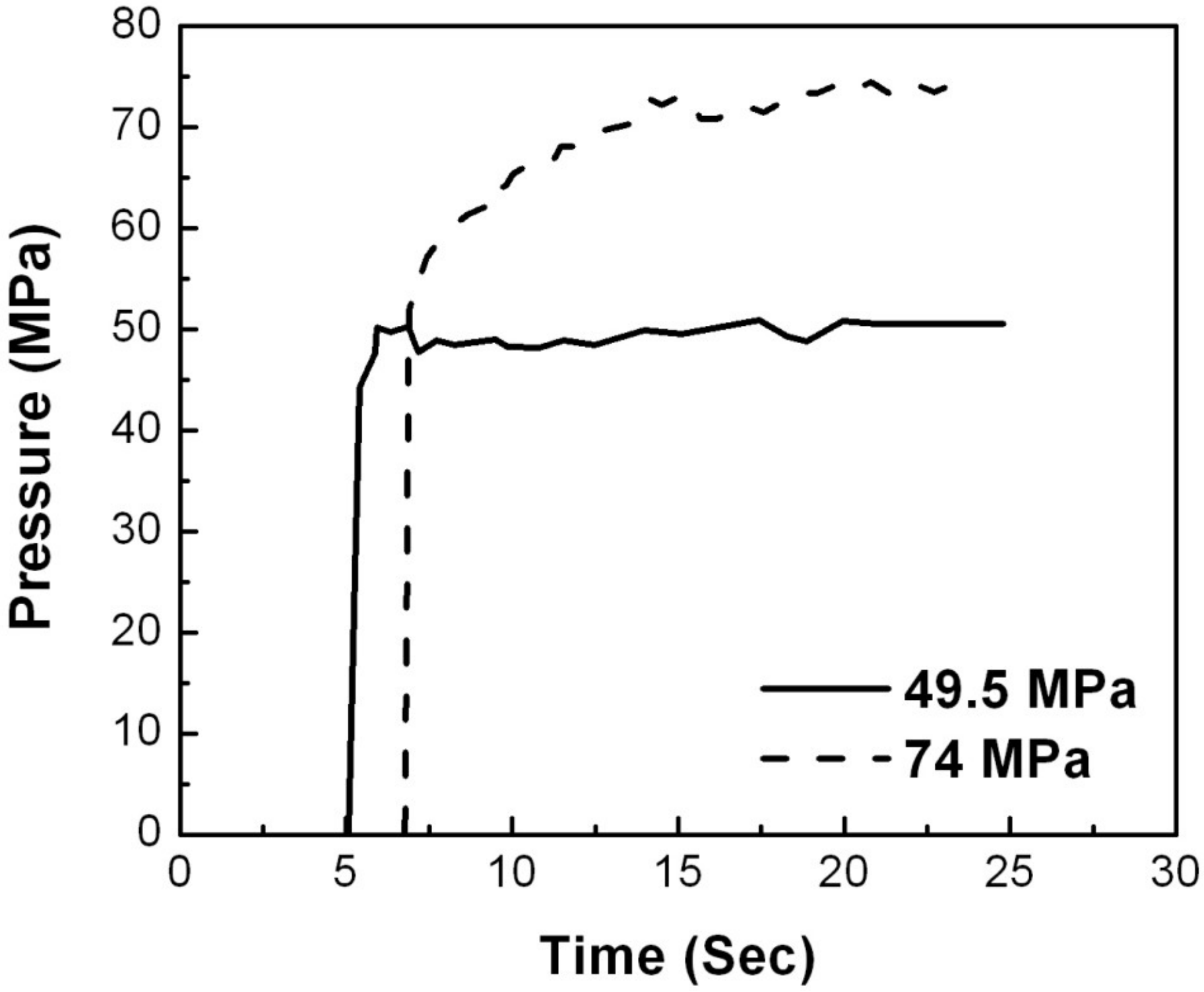
Pressure (MPa)
Time (Sec)
49.5 MPa
74 MPa
0
10
20
30
40
50
60
70
80
0
5
10
15
20
25
30

Figure 3

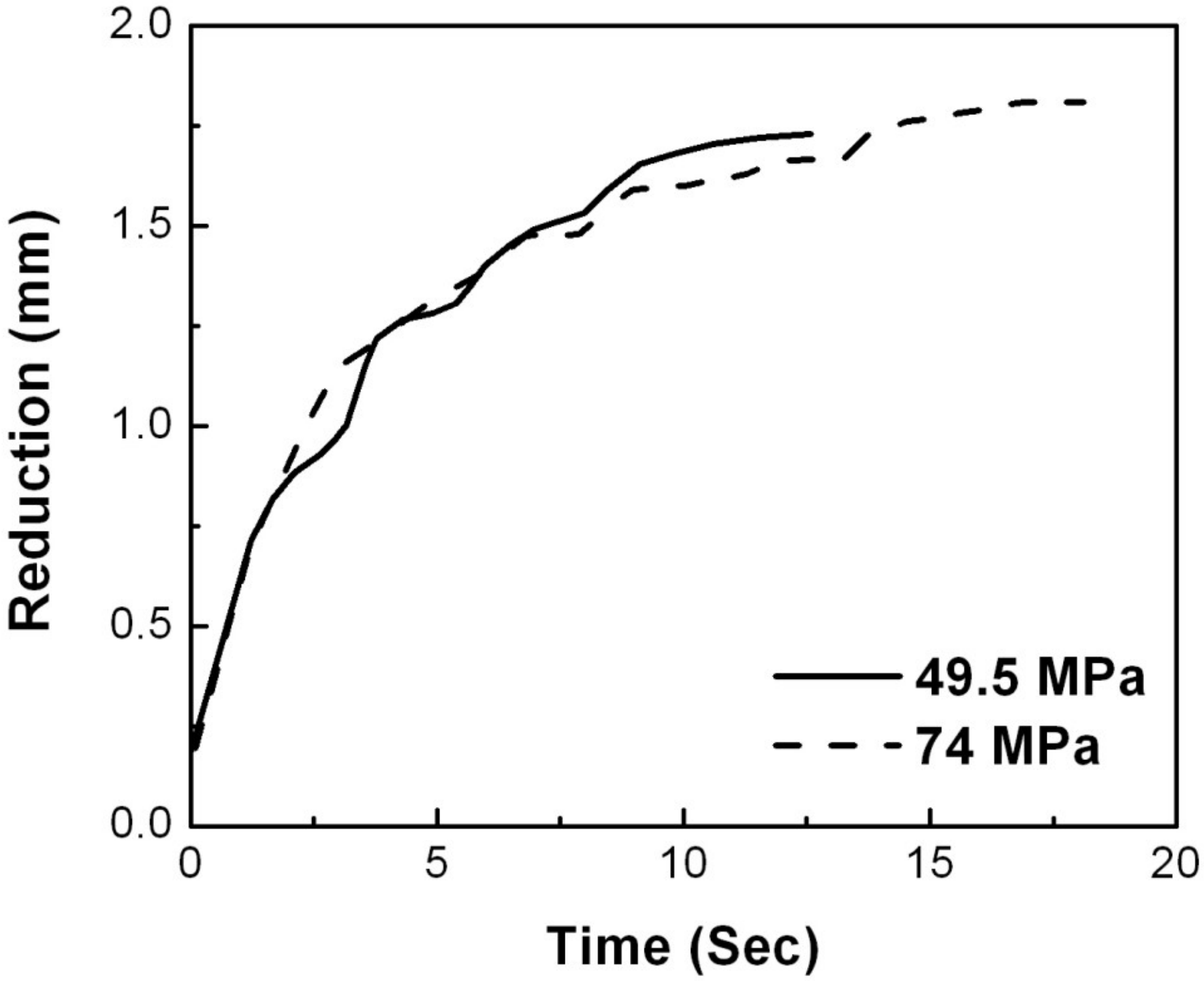

Figure 4

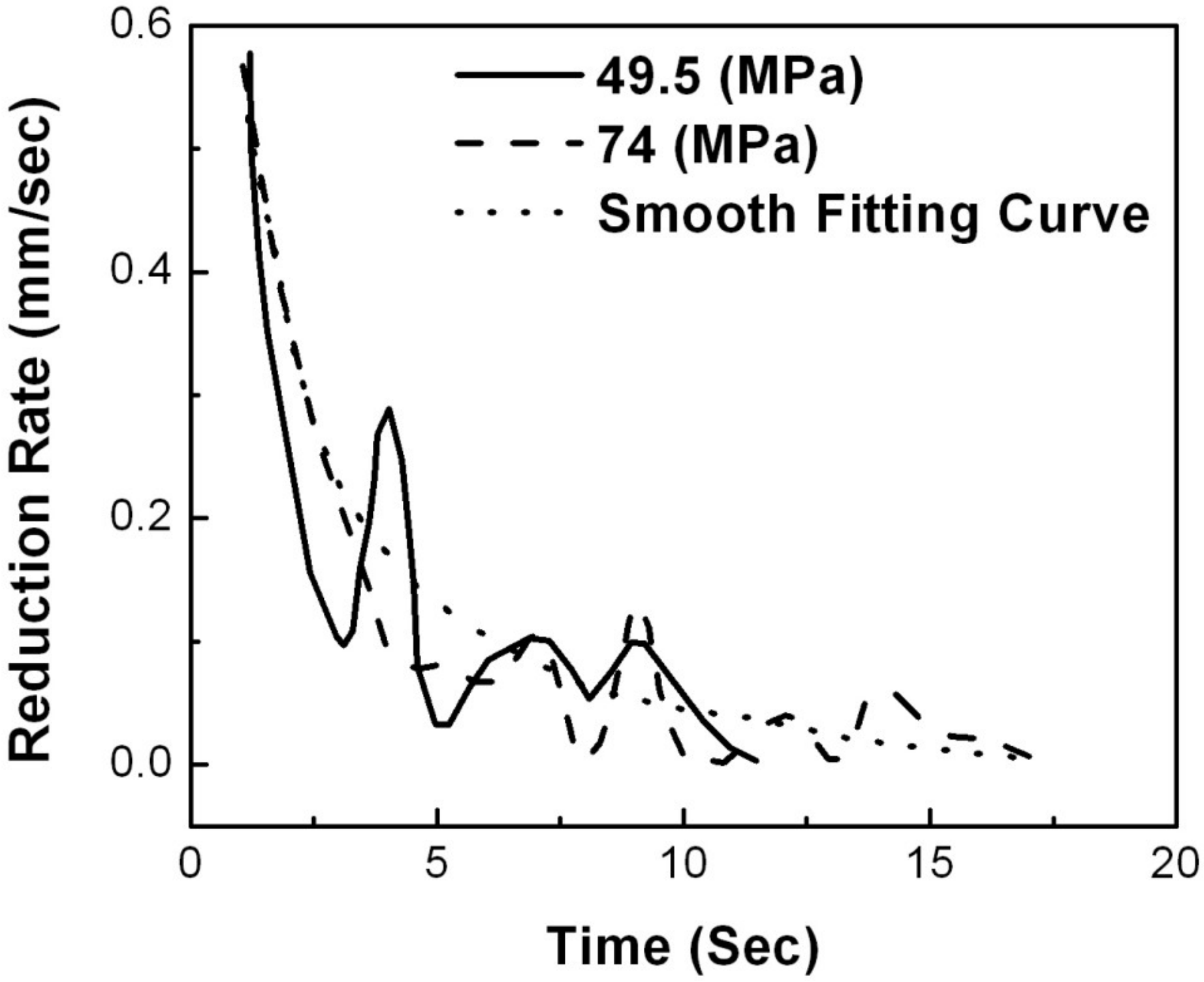

49.5 (MPa)
74 (MPa)
Smooth Fitting Curve
Reduction Rate (mm/sec)
Time (Sec)
0.6
0.4
0.2
0.0
0
5
10
15
20